\documentstyle[11pt,epsfig]{article}

\def\laq{~\raise 0.4ex\hbox{$<$}\kern -0.8em\lower 0.62
ex\hbox{$\sim$}~}
\def\gaq{~\raise 0.4ex\hbox{$>$}\kern -0.7em\lower 0.62
ex\hbox{$\sim$}~}

\begin{document}

\begin{titlepage}
\begin{flushright}
CERN-PH-TH/2010-281
\end{flushright}
\vspace*{1.5 cm}

\begin{center}
\huge{Reynolds numbers in the early Universe}
\vskip 1cm
\large{Massimo Giovannini\footnote{e-mail address: massimo.giovannini@cern.ch}}
\vskip 0.5cm
{\it   Department of Physics, Theory Division, CERN, 1211 Geneva 23, Switzerland}
\vskip 0.5cm
{\it  INFN, Section of Milan-Bicocca, 20126 Milan, Italy}
\vskip 1cm

\begin{abstract}
After electron-positron annihilation and prior to photon decoupling the magnetic Reynolds number is approximately twenty orders of magnitude larger than its kinetic counterpart which is, in turn, smaller than one. In this globally neutral system
the large-scale inhomogeneities are provided by the spatial fluctuations of the scalar curvature. Owing to the analogy with the description of Markovian conducting fluids in the presence of acoustic fluctuations, the evolution equations of a putative magnetic field are averaged over the large-scale flow determined by curvature perturbations. General lessons are drawn on the typical diffusion scale of magnetic inhomogeneities. It is speculated that Reynolds numbers prior to electron-positron annihilation can be related to the entropy contained in the Hubble volume during the various stages of the evolution of the conducting plasma.
\end{abstract}
\end{center}
\end{titlepage}

\newpage
In a conducting plasma, such as the early Universe, the kinetic and magnetic Reynolds numbers are defined as \cite{bisk,goed,mg1}
\begin{equation}
R_{\mathrm{kin}} = \frac{v_{\mathrm{rms}}\, L_{v}\, }{\nu_{\mathrm{th}}}, \qquad R_{\mathrm{magn}} = \frac{v_{\mathrm{rms}}\, L_{B}\, }{\nu_{\mathrm{magn}}},\qquad Pr_{\mathrm{magn}} = \frac{R_{\mathrm{magn}}}{R_{\mathrm{kin}}},
\label{R0} 
\end{equation}
where $v_{\mathrm{rms}}$ estimates the bulk velocity of the plasma while $\nu_{\mathrm{th}}$ and $\nu_{\mathrm{magn}}$ are the coefficients of thermal and magnetic diffusivity; $L_{v}$ and $L_{B}$ are, respectively, the correlation scales of the velocity field and of the magnetic field.   In Eq. (\ref{R0}) $Pr_{\mathrm{magn}}$ denotes the so-called magnetic Prandtl number  \cite{bisk,goed,mg1}. 

Prior to electron-positron annihilation (i.e. $T \geq \mathrm{MeV}$) the coefficient of thermal diffusivity can be estimated as 
 $\nu_{\mathrm{th}} \sim (\alpha_{\mathrm{em}}^2 T)^{-1}$ from the two-body scattering of relativistic species with significant momentum transfer. The conductivity 
of the plasma is $\sigma \sim T/\alpha_{\mathrm{em}}$ so that the magnetic diffusivity becomes $\nu_{\mathrm{magn}} = \alpha_{\mathrm{em}} (4\pi T)^{-1}$. Assuming, for sake of simplicity, thermal and kinetic equilibrium of all relativistic species (which is not exactly the case for $T\sim \mathrm{MeV}$) the kinetic Reynolds number turns out to be $R_{\mathrm{kin}} \simeq {\mathcal O}(10^{16})$, the magnetic Reynolds number 
 is $R_{\mathrm{magn}} \simeq 4\pi/\alpha_{\mathrm{em}}^3 R_{\mathrm{kin}} \sim {\mathcal O}(10^{24})$ and $Pr_{\mathrm{magn}} \sim 10^{7}$. The latter estimates have been obtained by assuming, in Eq. (\ref{R0}), 
 $L_{v} \simeq L_{B} \sim H^{-1}$ (where $H^{-1}$ is the Hubble radius at the corresponding 
 epoch); when the evolution of the 
 background geometry is decelerated (i.e. $a(t) \sim t^{\epsilon}$ with $0< \epsilon< 1$) =
 the particle horizon coincides with the Hubble radius up to an immaterial numerical factor 
 (i.e. $\epsilon/(1- \epsilon)$) which shall be neglected throughout\footnote{In what follows 
 we shall assume a conformally flat Friedmann-Robertson-Walker background geometry $\overline{g}_{\mu\nu} = a^2(\tau) \eta_{\mu\nu}$ where $\eta_{\mu\nu}$ is the Minkowski metric, $\tau$ the conformal time coordinate and $a(\tau)$ the 
 scale factor; the conformal time coordinate $\tau$ is related to the cosmic time $t$ as $a(\tau) d\tau = dt$.}. 
 In the symmetric phase of the standard electroweak theory, picking up a temperature $T > 100 \, \mathrm{GeV}$  where all the species (including the Higgs boson and the top quark) are in thermal and kinetic equilibrium, 
 $\nu_{\mathrm{th}}$ and $\nu_{\mathrm{magn}}$ can be computed \cite{mg2} in terms of the hypercharge coupling constant since the non-screened vector modes at finite conductivity are associated with the hypercharge field. In the electroweak case $R_{\mathrm{kin}} \sim {\mathcal O}(10^{11})$ and $R_{\mathrm{magn}} \sim {\mathcal O}(10^{17})$.

The hypothesis of primeval turbulence has been a recurrent theme since the first speculations 
on the origin of the light nuclear elements. The implications of turbulence for galaxy formation 
have been pointed out in the fifties by Von Weizs\"aker and Gamow \cite{VW}. They have been scrutinized in the sixties and early seventies by various authors \cite{turb1} (see also \cite{peebles,barrow} and discussions therein). In the eighties it has been argued \cite{hogan} that first-order phase transitions in the early Universe, if present,  can provide a source of kinetic turbulence and, hopefully, the possibility of inverse cascades which could lead to an enhancement of the correlation scale of a putative large-scale magnetic field, as discussed in
in \cite{olesen1,olesen2} (see also \cite{enqvist} and references therein). The limits $R_{\mathrm{kin}} \gg 1$ and $R_{\mathrm{magn}} \gg 1$  are customarily assumed in the scrutiny of hydromagnetic turbulence where both the magnetic flux and the magnetic helicity are conserved since $\nu_{\mathrm{magn}} \ll 1$ i.e.\footnote{In hydromagnetic turbulence it is customarily assumed that $Pr_{\mathrm{magn}} \simeq 1$ while 
the flow is incompressible (i.e. the bulk velocity of the plasma is solenoidal) \cite{bisk,goed}. This is not necessarily the case in the early Universe, as we shall see.}
 \begin{equation}
\frac{d}{d\tau} \int_{\Sigma} \vec{B} \cdot d\vec{\Sigma}=- \nu_{\mathrm{magn}} \int_{\Sigma} \vec{\nabla} \times(\vec{\nabla}
\times\vec{B})\cdot d\vec{\Sigma},\qquad \frac{d}{d \tau}\int_{V} d^3 x \vec{A}~\cdot \vec{B} = - 2 \nu_{\mathrm{magn}} \int_{V} d^3 x
{}~\vec{B}(\cdot\vec{\nabla} \times\vec{B}),
\label{FC}
\end{equation}
where $V$ and $\Sigma$ are a fiducial volume and a fiducial surface moving with the conducting fluid; $\vec{B}$ and $\vec{A}$ 
denote the comoving magnetic field and the comoving vector potential. In the 
ideal hydromagnetic limit (i.e. $\sigma \to \infty$, $\nu_{\mathrm{magn}} \to 0$ and $R_{\mathrm{magn}} \to \infty$)
the flux is exactly conserved and the number of links and twists in the magnetic flux lines is also preserved by the time evolution.  If $R_{\mathrm{kin}} \gg 1$ and  $R_{\mathrm{magn}} \leq {\mathcal O}(1)$  the system is still turbulent; however, since the total time derivative of the magnetic flux and of the magnetic helicity are both ${\mathcal O}(\nu_{\mathrm{magn}})$ the terms at the right hand side of Eq. (\ref{FC}) cannot be neglected.  Finally,  
if $R_{\mathrm{magn}} \gg 1$  and $R_{\mathrm{kin}} \ll 1$ the fluid is not kinetically turbulent but but the magnetic flux 
is conserved.

For $T> \mathrm{Mev}$ the Reynolds numbers can be viewed as a measure the entropy stored in a given Hubble volume.  The entropy stored within the Hubble volume $V_{H} \sim 4\pi H^{-3}(t)/3$ is
directly expressible in terms of the Reynolds numbers:
\begin{equation}
 S_{H} = \frac{4}{3} \pi s H^{-3} = \frac{8 \pi^3}{135} \, \frac{N_{\mathrm{eff}}}{v_{\mathrm{rms}}^3\, 
 \alpha_{\mathrm{em}}^{6}} \, R_{\mathrm{kin}}^3,
 \label{R1}
 \end{equation}
 where $s$ is the entropy density of the plasma. The adiabatic expansion implies that the total entropy over a comoving volume is conserved. Conversely $S_{H}$, i.e. the entropy stored
in the Hubble volume,  increases as the plasma cools down since the temperature redshifts as 
$a^{-1}$ (where $a$ is the scale factor)  but the Hubble volume typically increases faster than $a^{3}$ both during radiation (i.e. $V_{H} \sim a^{6}$) and during matter (i.e. $V_{H} \sim a^{9/2}$).  Up to numerical factors $S_{H} \sim {\mathcal O}(10^{64})$ right before electron-positron annihilation and 
$S_{H}  \sim {\mathcal O}(10^{48})$ in the symmetric phase of the electroweak theory. 
The maximal entropy stored in the Hubble volume today\footnote{We do not consider here the possibility of a gravitational entropy associated with a Hubble screen approximately saturating the Hawking-Bekenstein bound and  implying $S_{\mathrm{screen}} \sim H_{0}^{-2} M_{\mathrm{P}}^2 \simeq {\mathcal O}(10^{120}) \gg S_{\gamma}$.} is obtained by integrating the entropy density of the Cosmic Microwave Background radiation (CMB in what follows) over the present value of the Hubble volume:
\begin{equation}
S_{\gamma} = \frac{4}{3} \pi s_{\gamma} H_{0}^{-3} \simeq 
1.43 \times 10^{88} \,\, \biggl(\frac{h_{0}}{0.7}\biggr)^{-3}, \qquad s_{\gamma}  = 
\frac{4}{45} \pi^2 T_{\gamma}^3,
\label{Sg}
\end{equation}
where $T_{\gamma} =2.725\,\, \mathrm{K}$. In the standard lore the huge value of the entropy contained in the Hubble volume is the result of an appropriate theory of  the initial conditions, since the adiabaticity condition can only 
be mildly violated after inflation and for a standard thermal history.

According to Eqs. (\ref{R1})--(\ref{Sg}) it would be tempting to establish a causal connection between the 
Hubble entropy of the CMB and the largeness  of the Reynolds numbers. While such a connection cannot 
be excluded for $T > \mathrm{MeV}$, 
Eqs. (\ref{R1}) and (\ref{Sg}), taken at face value,  would imply that  the kinetic and magnetic Reynolds numbers must reach their  maximum for $T \ll \mathrm{MeV}$. The latter conclusion is incorrect insofar as the thermal 
diffusion coefficient sharply increases after electron-positron annihilation while 
the conductivity is only suppressed as $\sqrt{\overline{T}/(m_{\mathrm{e}} a)}$ where $\overline{T}= a \,T$ is the comoving 
temperature and $m_{\mathrm{e}}$ is the electron mass.  While prior to electron-positron annihilation $R_{\mathrm{kin}} \gg 1$ and $R_{\mathrm{magn}} \gg 1$, after $e^{+}$-$e^{-}$ annihilation $R_{\mathrm{magn}}$ is still very large,
$R_{\mathrm{kin}}$ gets smaller than $1$ and $Pr_{\mathrm{magn}}$ sharply increases. Indeed, prior to last scattering,  the thermal diffusivity is dominated by Thomson scattering and the concentration of the charge carriers is not of the order  of the photon concentration (as for $T> \mathrm{MeV}$) but ten orders of magnitude smaller i.e.  $n_{\mathrm{e}} = \eta_{\mathrm{b}} n_{\gamma}$ where $n_{\gamma}$ is the comoving concentration of the photons and $\eta_{\mathrm{b}} = {\mathcal O}(10^{-10})$ is the ratio of the baryonic concentration to the photon concentration indirectly probed by big-bang nucleosynthesis 
and affecting the present abundances of light nuclear elements. The thermal diffusion coefficient is then given by:
\begin{equation}
\nu_{\mathrm{th}}(\tau) = \frac{4}{5} c_{\mathrm{s\, b}}^2(\tau) \, \lambda_{\mathrm{e}\gamma}(\tau), \qquad c_{\mathrm{s\, b}}(\tau) = \frac{1}{\sqrt{3 [ 1 + R_{\mathrm{b}}(\tau)]}},
\label{R2}
\end{equation}
where the electron-photon mean free path $\lambda_{\gamma\mathrm{e}}$ and the ratio between 
the baryonic matter density and the photon energy density $R_{\mathrm{b}}$ are defined as:
\begin{equation}
\lambda_{\gamma\mathrm{e}} = \frac{a_{0}}{\tilde{n}_{\mathrm{e}} a(\tau) \sigma_{\gamma\mathrm{e}}}, \qquad R_{\mathrm{b}}(\tau) = \frac{3}{4} \biggl(\frac{\omega_{\mathrm{b}0}}{\omega_{\gamma 0}}\biggr) \biggl(\frac{a}{a_{0}}\biggr)
= \frac{685.62}{z+1} \biggl(\frac{\omega_{\mathrm{b}0}}{0.02258}\biggr); 
\label{R3}
\end{equation}
the baryon matter density $\rho_{\mathrm{b}}$ is the sum of the matter densities and of the ions and electrons; $\sigma_{\gamma \mathrm{e}}$ is the electron-photon cross section; $\omega_{\mathrm{b}0} = h_{0}^2 \Omega_{\mathrm{b}0}$ and $\Omega_{\mathrm{b}0}$ is the critical fraction of baryons.  For the sake of simplicity we shall adopt, in the explicit estimates, the following fiducial set of parameters 
\begin{equation}
( \Omega_{\mathrm{b}0}, \, \Omega_{\mathrm{c}0}, \Omega_{\mathrm{de}0},\, h_{0},\,n_{\mathrm{s}},\, \epsilon_{\mathrm{re}}) \equiv (0.0449,\, 0.222,\, 0.734,\,0.710,\, 0.963,\,0.088),
\label{R3a}
\end{equation}
which are determined from the WMAP 7yr data alone \cite{wmap7a} in the light of the vanilla $\Lambda$CDM scenario.  After electron-positron annihilation the conductivity given by binary collisions can be estimated as 
\begin{equation}
\sigma(\tau) = \sigma_{1}\frac{\overline{T}}{\alpha_{\mathrm{em}}}  \sqrt{\frac{\overline{T}}{m_{\mathrm{e}} a}} 
\frac{1}{\ln{\Lambda_{\mathrm{C}}}}, \qquad
 \Lambda_{\mathrm{C}}(\overline{T}) = \frac{3}{2 e^3} \biggl(\frac{\overline{T}^3}{\pi n_{\mathrm{e}}}\biggr)^{1/2} =  1.105\times 10^{8} \biggl(\frac{\omega_{\mathrm{b}0}}{0.02258}\biggr)^{-1/2},
 \label{R5}
\end{equation}
where $\sigma_{1} = 9/(8 \pi \sqrt{3})$ depends on the way multiple scattering is estimated and $\Lambda_{\mathrm{C}}$ is the argument of the Coulomb logarithm.  

Let us finally come to a more detailed estimate of the velocity field prior to last scattering. 
The bulk velocity is defined as the center of mass velocity 
of the positive and negative charge carriers present in the globally neutral plasma\footnote{If the charge carriers coincide with electrons and ions, denoting with $m_{\mathrm{i}}$ and $m_{\mathrm{e}}$ the masses of the electrons and ions
the bulk velocity of the plasma is defined as $\vec{v}_{\mathrm{b}} =  (m_{\mathrm{e}} \vec{v}_{\mathrm{e}} + 
m_{\mathrm{i}} \vec{v}_{\mathrm{i}})/(m_{\mathrm{e}} + m_{\mathrm{i}})$. The center of mass velocity of the electron-ion system is often called baryon velocity.}. 
The customary assumptions of hydromagnetic turbulence imply a solenoidal bulk velocity 
field with $Pr_{\mathrm{magn}} \simeq {\mathcal O}(1)$ \cite{bisk,goed}. Conversely,
prior to photon decoupling, the bulk velocity of the plasma is not solenoidal (i.e. $\vec{\nabla}\cdot\vec{v}_{\mathrm{b}} \neq 0$) and $Pr_{\mathrm{magn}} \gg 1$. This means that the plasma is compressible and the divergence of the bulk velocity 
is directly affected by the large-scale curvature fluctuations.
Using Eq. (\ref{R3}) and assuming  $v_{\mathrm{rms}} =1$  the kinetic and the magnetic Reynolds 
numbers can be estimated as $R_{\mathrm{kin}} \simeq 0.03$ and $R_{\mathrm{magn}} \simeq 1.30 \times 10^{20}$ for a typical last-scattering redshift $z\sim 1090$.  The upper limit obtained on $R_{\mathrm{kin}}$ 
by assuming that $v_{\mathrm{rms}}$ coincides with the speed of light can be made more stringent 
since the large-scale flow, prior to last scattering, can be determined from the evolution equations of the baryon-photon system.  

For $T< \mathrm{MeV}$ and around last-scattering the differences between the velocities of the baryons and of the photons
 are quickly washed out because of the tight-coupling between ions, electrons and photons. Recalling that ${\mathcal H} =\partial_{\tau} \ln{a} = a H$,  to lowest order in the tight-coupling approximation the truncated set of hydromagnetic equations reads \cite{mg3}
\begin{eqnarray}
&& \partial_{\tau} \vec{v}_{\gamma\mathrm{b}} + \frac{{\mathcal H} R_{\mathrm{b}}}{R_{\mathrm{b}} + 1} \vec{v}_{\gamma\mathrm{b}} = \frac{R_{\mathrm{b}}}{R_{\mathrm{b}} + 1} \frac{\vec{J} \times \vec{B}}{\rho_{\mathrm{b}} a^4} - 
\frac{\vec{\nabla} \delta_{\gamma}}{4 (R_{\mathrm{b}} + 1)} - \vec{\nabla} \phi  + \nu_{\mathrm{th}} \nabla^2 \vec{v}_{\gamma\mathrm{b}},
\label{pd2}\\
&& \partial_{\tau} \delta_{\mathrm{b}} = 3 \partial_{\tau} \psi - \vec{\nabla}\cdot \vec{v}_{\gamma\mathrm{b}} + \frac{\vec{J} \cdot \vec{E}}{\rho_{\mathrm{b}} a^4},\qquad \partial_{\tau} \delta_{\gamma} = 4 \partial_{\tau}\psi - \frac{4}{3} \vec{\nabla} \cdot\vec{v}_{\gamma\mathrm{b}},
\label{pd4}
\nonumber\\
&& \partial_{\tau} \vec{B} = \vec{\nabla}\times (\vec{v}_{\gamma\mathrm{b}} \times \vec{B}) + \nu_{\mathrm{magn}} \nabla^2 \vec{B}
+ \vec{\nabla}\times \biggl( \frac{\vec{\nabla} p_{\mathrm{e}}}{e n_{\mathrm{e}}} \biggr) - \frac{1}{4\pi e n_{\mathrm{e}}}  \vec{\nabla}\times[
(\vec{\nabla} \times \vec{B})\times \vec{B}],
\label{pd5}
\end{eqnarray}
where $\delta_{\mathrm{b}}$ and $\delta_{\gamma}$ denote the density contrasts of the baryons and of the 
photons in the longitudinal gauge; $\phi$ and $\psi$ denote, respectively, the $(00)$ and $(ii)$ fluctuations 
of the conformally flat background geometry adopted in the present discussion. The 
total Ohmic current $\vec{J}$ obeys an evolution equation which can be reduced to a consistency condition 
as in the Eckart approach to relativistic thermodynamics; such a relation is given by:
\begin{equation}
\vec{J} = \sigma\biggl( \vec{E} + \vec{v}_{\gamma\mathrm{b}} \times \vec{B} + \frac{\vec{\nabla} p_{\mathrm{e}}}{e n_{\mathrm{e}}} - 
\frac{\vec{J} \times \vec{B}}{e n_{\mathrm{e}}} \biggr),
\label{pd5a}
\end{equation}
where $n_{\mathrm{e}} = a^3 \tilde{n}_{\mathrm{e}}$. The system of Eqs. (\ref{pd2})--(\ref{pd5a}) 
is supplemented by the evolution equations of the curvature perturbations which have 
been written elsewhere in the longitudinal gauge and even in full gauge-invariant terms (see \cite{mg3} and references therein). 

In previous studies \cite{mg1,mg3} the emphasis has been to see which are the effects of the large-scale magnetic fields 
on the scalar modes of the geometry. The hierarchy between $R_{\mathrm{kin}}$ and $R_{\mathrm{magn}}$ 
after electron-positron annihilation suggests the possibility  of addressing also 
the complementary part of the problem, i.e. the effect of the large-scale flow on the evolution of the magnetic field. 
Neglecting the terms which are quadratic in the magnetic field as well as 
the thermoelectric term (i.e. the last two contributions in Eq. (\ref{pd5a})) the large-scale flow 
can be determined by using a standard WKB analysis giving, in Fourier space,
\begin{eqnarray}
\vec{v}_{\gamma\mathrm{b}}(k,\tau) &=& i \, \hat{k} \, \,\overline{M}_{\mathcal R}(k,\tau)  \sin{[k r_{\mathrm{s}}(\tau)]} \, e^{- k^2/k_{\mathrm{d}}^2}, \qquad \hat{k} = \frac{\vec{k}}{k},
\nonumber\\
r_{\mathrm{s}}(\tau) &=& \int_{0}^{\tau}
  c_{\mathrm{sb}}(\tau') \, d\tau',\qquad \frac{1}{k_{\mathrm{d}}^2(\tau)} = \frac{1}{2} \,\,
  \int_{0}^{\tau} \,\,\nu_{\mathrm{th}}(\tau')\, \, d\tau',
\label{IM1}
\end{eqnarray}
where $r_{\mathrm{s}}(\tau)$ and $k_{\mathrm{d}}(\tau)$ denote, respectively, the sound horizon and the typical scale 
of diffusive damping; $\overline{M}_{\mathcal R}(k,\tau)$ encodes the normalization 
inherited from the (adiabatic) curvature perturbations which are the only source of large-scale inhomogeneities 
in the vanilla $\Lambda$CDM scenario:
\begin{equation}
\overline{M}_{{\mathcal R}}(k,\tau) = 3^{1/4}\, c^{3/2}_{\mathrm{sb}}(\tau) \,\biggl[ \frac{1}{c_{\mathrm{sb}}^2(\tau)} - 2\bigg] T_{{\mathcal R}}(\tau) {\mathcal R}_{*}(\vec{k}),\qquad T_{{\mathcal R}}(\tau)  = 1 - \frac{{\mathcal H}}{a^2} \int_{0}^{\tau} a^2(\tau') d\tau'.
\label{IM2}
\end{equation}
As a result of Eqs. (\ref{pd2})--(\ref{pd5}), the velocity field (\ref{IM1}) is not solenoidal. 
This is situation differs from standard hydromagnetic turbulence 
and it is closer to the situation of acoustic turbulence (see e.g. \cite{vain1}, first paper). 
Following the standard conventions \cite{wmap7a},
the correlation function of curvature perturbations in Fourier space is:
\begin{equation}
\langle {\mathcal R}_{*}(\vec{p}) {\mathcal R}_{*}(\vec{q}) \rangle = \frac{2\pi^2}{q^3} \delta^{(3)}(\vec{q} + \vec{p}) 
{\mathcal P}_{{\mathcal R}}(q),\qquad {\mathcal P}_{{\mathcal R}}(q) = {\mathcal A}_{{\mathcal R}} \biggl(\frac{q}{q_{\mathrm{p}}}\biggr)^{n_{\mathrm{s}}-1},
\label{IM7}
\end{equation}
where, according to Eq. (\ref{R3a}), $n_{\mathrm{s}} =0.963$ and ${\mathcal A}_{{\mathcal R}} = (2.43\pm 0.11)\times 10^{-9}$;
$q_{\mathrm{p}} =0.002\, \mathrm{Mpc}^{-1}$ denotes the pivot scale at which the power spectrum of curvature 
perturbations is conventionally normalized.
Given the intrinsic inhomogeneity of the large-scale velocity flow, it is natural to
generalize the kinetic and the magnetic Reynolds numbers of Eq. (\ref{R0}) to Fourier space by keeping the dependence on the wavenumbers in the velocity, the dependence on the redshift in the diffusion coefficients and by choosing $L_{v} \simeq L_{B} = 1/k_{\mathrm{phys}}(\tau)$ where $k_{\mathrm{phys}} = k (z +1)$: 
\begin{eqnarray}
R_{\mathrm{kin}}(k,z) &=& {\mathcal N}_{\mathrm{kin}}\,\, F(k,z),\qquad Pr_{\mathrm{magn}}(k,z) = {\mathcal N}_{\mathrm{magn}}
\,\, G(k,z) 
\nonumber\\
{\mathcal N}_{\mathrm{kin}} &=& \frac{15\, \pi}{2\,3^{3/4}} \frac{\alpha_{\mathrm{em}}^2}{m_{\mathrm{e}}^2 H_{0}} \eta_{\mathrm{b}} n_{\gamma} \sqrt{{\mathcal A}_{\mathcal R}}, \qquad {\mathcal N}_{\mathrm{magn}} = \frac{9 \pi^2}{20 \zeta(3) \alpha_{\mathrm{em}}} \frac{\sigma_{1}}{\ln{\Lambda_{\mathrm{C}}(\overline{T})}\,\eta_{\mathrm{b}}} \biggl(\frac{m_{\mathrm{e}}}{\overline{T}}\biggr)^{3/2}, 
\nonumber\\
F(k, z) &=& H_{0}\, [1 - 2c_{\mathrm{sb}}^2(z)]  T_{\mathcal R}(z,z_{\mathrm{eq}}) \biggl(\frac{k}{k_{\mathrm{p}}}\biggr)^{n_{\mathrm{s}} -1} \frac{\sin{[k\,r_{\mathrm{s}}(z)]}}{k\, c^{5/2}_{\mathrm{sb}}(z)},\qquad G(k,z) = \frac{c_{\mathrm{sb}}^2(z)}{(z + 1)^{3/2}}, 
\label{R11}
\end{eqnarray}
where $\zeta(3)= 1.202$; introducing the function $y(z,z_{\mathrm{eq}}) = [\sqrt{1 +( z_{\mathrm{eq}}+1)/(z+1)} -1]$ 
the function $T_{\mathcal R}(z,z_{\mathrm{eq}})$ can be expressed as:
\begin{equation}
T_{\mathcal R}(z,z_{\mathrm{eq}})  = (z + 1) \biggl\{1 - \frac{2}{15}\frac{[y(z,z_{\mathrm{eq}}) +1 ] [ 3 y^2(z,z_{\mathrm{eq}}) + 15 
y(z,z_{\mathrm{eq}}) + 20]}{[2 + y(z,z_{\mathrm{eq}})]^3}\biggr\} .
\label{R12}
\end{equation}
\begin{figure}[!ht]
\centering
\includegraphics[height=5.1cm]{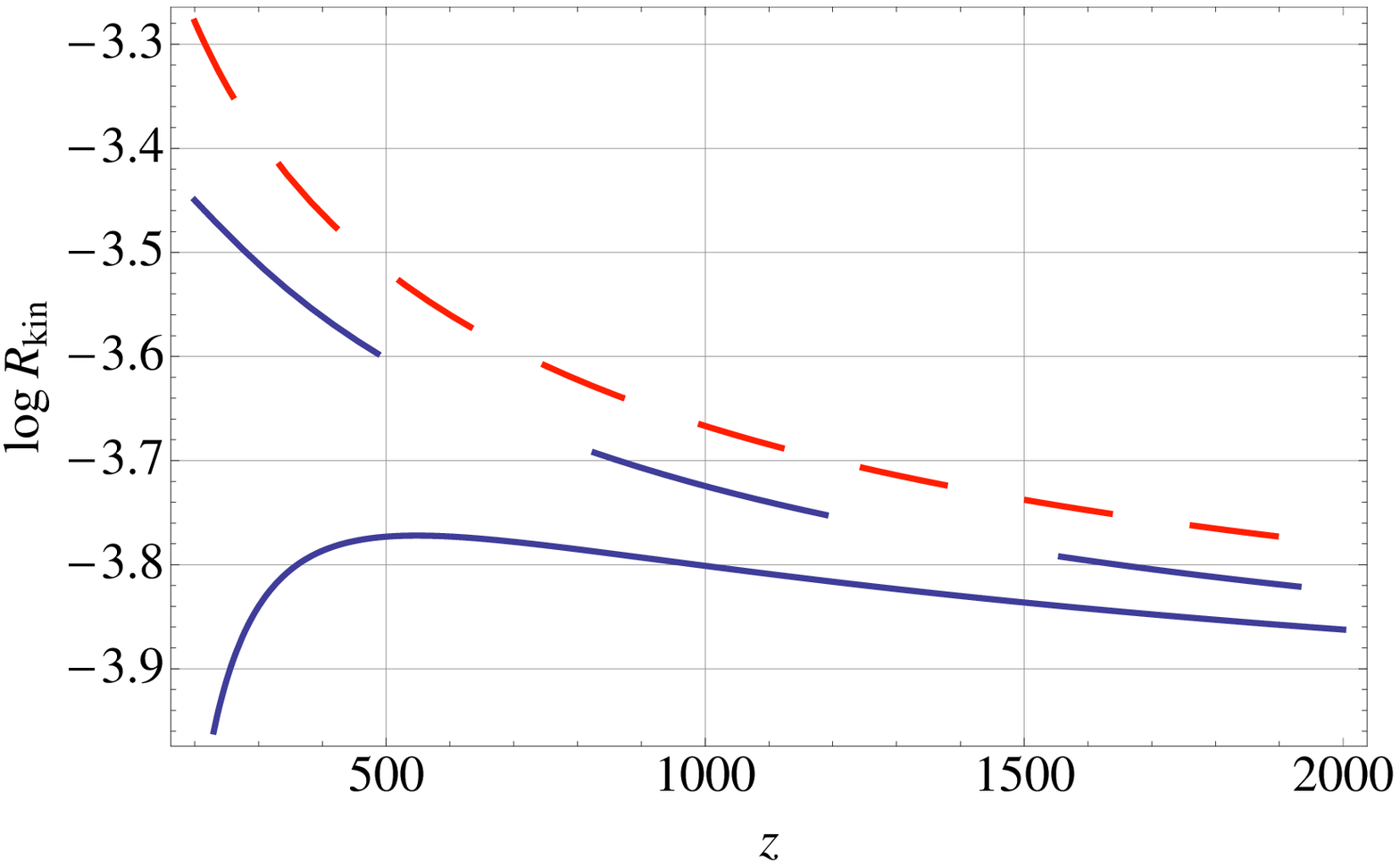}
\includegraphics[height=5.1cm]{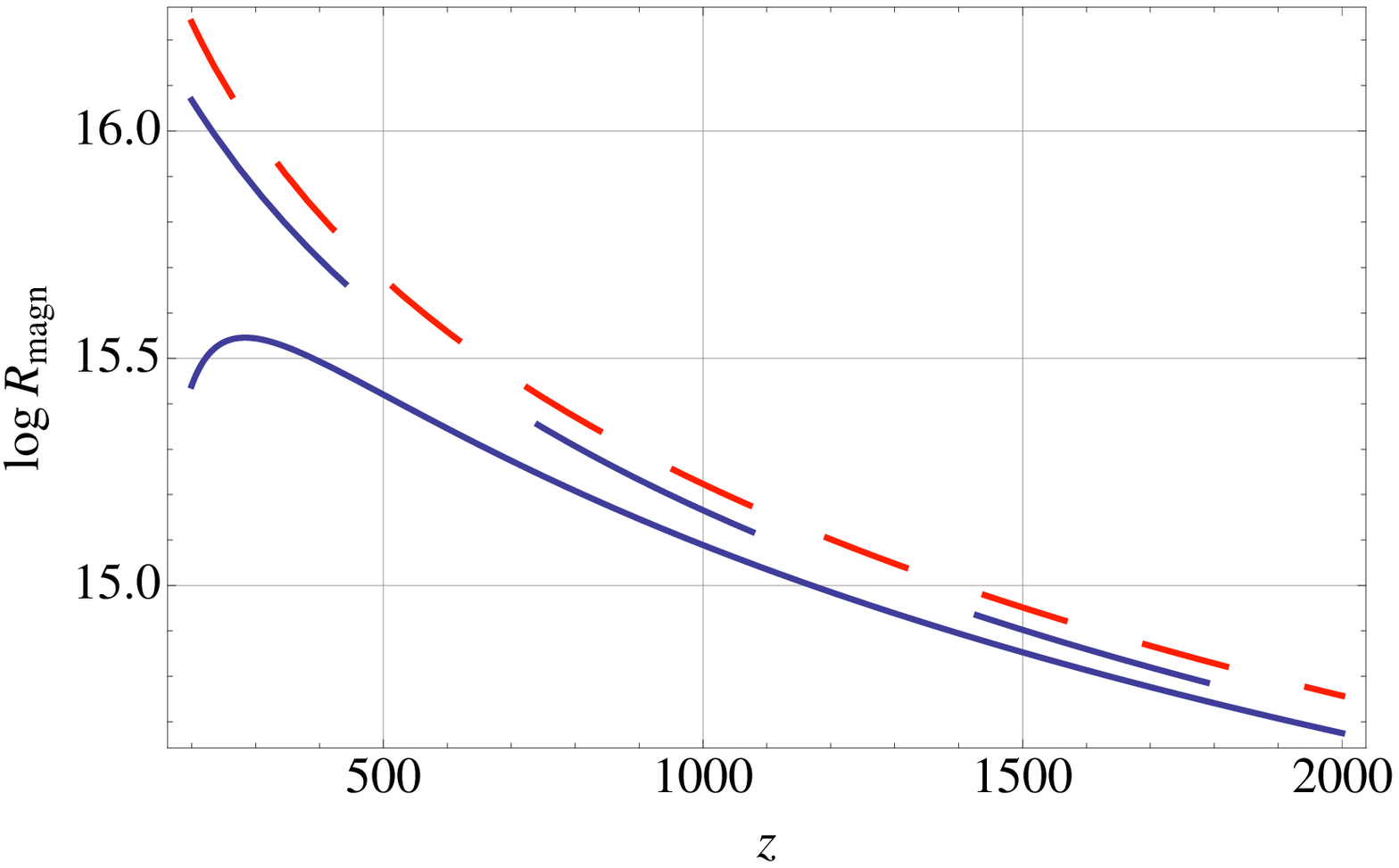}
\caption[a]{The kinetic Reynolds number (left plot) and the magnetic Reynolds number (right plot)  for $ k = 0.0002\, \mathrm{Mpc}^{-1}$ (short dashed line),  
$ k = 0.002\, \mathrm{Mpc}^{-1}$ (long dashed line) and $ k = 0.004\, \mathrm{Mpc}^{-1}$ (full line). In both plots on the vertical axis the common logarithm of the corresponding quantity is illustrated.}
\label{F1}      
\end{figure}
In Fig. \ref{F1} the kinetic and the magnetic Reynolds numbers are illustrated for the set of cosmological parameters of Eq. (\ref{R3a}). The three curves in each plot correspond to three different wavenumbers. For even larger 
wavenumbers both quantities are oscillating as it can be argued from Eq. (\ref{R11}).

The hierarchy between the kinetic and the magnetic Reynolds number defines naturally a  
perturbative scheme where the evolution equations of the magnetic field can be averaged over the 
large-scale flow. Consider the magnetic diffusivity equation (\ref{pd5})  and neglect all the terms which are of higher order in the magnetic field intensity. An iterative solution of Eq. (\ref{pd5}) can then be constructed as 
\begin{eqnarray}
 B_{i}(\vec{k},\tau) &=& \sum_{n =0}^{\infty}  B^{(n)}_{i}(\vec{k},\tau),\qquad {\mathcal G}_{k}(y) = 
 e^{- k^2 \nu_{\mathrm{magn}}\,y}
\label{NM14aa}\\
B^{(n+1)}_{i}(\vec{k},\tau) &=& \frac{(-i)}{(2\pi)^{3/2}} \, \int_{0}^{\tau} {\mathcal G}_{k}(\tau-\tau_{1}) \, d\tau_{1} \, \int d^{3} q \, \int d^3 p  \,\,\, \delta^{(3)}(\vec{k} - \vec{q} - \vec{p})\,
\nonumber\\
&\times& \epsilon_{m\, n\, i} \, \epsilon_{a\, b\, n} \, (q_{m} + p_{m}) \, v_{a}(\vec{q},\tau_{1}) \, B^{(n)}_{b}(\vec{p},\tau_{1}),
\label{NM14}
\end{eqnarray}
where, for simplicity, $\nu_{\mathrm{magn}}$ is assumed to be constant in time. From Eq. (\ref{NM14}) the first few terms of the recursion are $B_{i}^{(0)}(\vec{k},\tau)$, $B_{i}^{(1)}(\vec{k},\tau)$ and $B_{i}^{(2)}(\vec{k},\tau)$. The term $B_{i}^{(0)}(\vec{k},\tau) = {\mathcal G}_{k}(\tau) B_{i}(\vec{k})$ where $B_{i}(\vec{k})$ parametrizes the initial stochastic magnetic field obeying flux conservation (see Eq. (\ref{FC})):
\begin{equation}
\langle B_{i}(\vec{k}) \, B_{j}(\vec{k}^{\,\,\prime}) \rangle = \frac{2\pi^2}{k^3} P_{ij}(\hat{k}) P_{B}(k) \, \delta^{(3)}(\vec{k} + \vec{k}^{\,\,\prime}),\qquad P_{ij}(\hat{k}) = \delta_{i j} - \hat{k}_{i} \, \hat{k}_{j}.
\label{NM15}
\end{equation}
Since the curvature perturbations are distributed as in Eq. (\ref{IM7}), the correlation 
function of the velocity for unequal times can be written as
\begin{equation}
\langle v_{i}(\vec{q},\tau) \, v_{j}(\vec{p},\tau^{\,\prime})\rangle = \frac{q_{i} \, q_{j}}{q^2} \,{\mathcal U}(q, |\tau - \tau^{\,\prime}|)\, \delta^{(3)}(\vec{q} + \vec{p}) , 
\qquad 
{\mathcal U}(q, |\tau - \tau^{\,\prime}|)= v(q) \, \tau_{\mathrm{c}} \, \delta(\tau - \tau^{\,\prime}),
\label{IM7A}
\end{equation}
where, to avoid confusions with vector indices, the subscript $\gamma\mathrm{b}$  has been suppressed.
The function $v(q)$ appearing in Eq. (\ref{IM7A}) is 
\begin{equation}
v(q) = \tau_{\mathrm {c}} {\mathcal V}(q), \qquad {\mathcal V}(q) = \overline{M}^2_{{\mathcal R}}(q,\tau_{*}) \frac{2\pi^2}{q^3}\, {\mathcal P}_{\mathcal R}(q)\, \sin^2{[q r_{\mathrm{s}}(\tau_{*})] } e^{ - 2 q^2/q_{\mathrm{d}}^2},
\label{IM7B}
\end{equation}
where $\tau_{*}$ denotes the last-scattering time and 
the correlation time $\tau_{\mathrm{c}}$ is the smallest time-scale when compared with other characteristic times arising in the problem. Because of the exponential suppression of the velocity correlation 
function for $\tau > \tau_{\mathrm{d}}$ (where $\tau_{\mathrm{d}}$ denotes the Silk time \cite{silk}), $\tau_{\mathrm{c}}$ approximately coincides with $\tau_{\mathrm{d}}$. The form of the correlator given in Eq. (\ref{IM7A}) is characteristic of Markovian conducting fluids \cite{vain1,vain2}.

Denoting with  $H_{i}(\vec{k},\tau)$  the magnetic field averaged over the fluid flow, 
the terms containing an odd number of velocities will be zero while the correlators containing an even number of velocities do not vanish
i.e. $ \langle B_{i}^{(2 n + 1)} \rangle =  H_{i}^{(2 n + 1)} =0$  and $\langle B_{i}^{(2 n + 2)} \rangle =  H_{i}^{(2 n + 2)} \neq 0$.
So, for instance, $\langle B_{i}^{(1)}\rangle =0$ while  $\langle B_{i}^{(2)} \rangle =  H_{i}^{(2)}$ is  
\begin{eqnarray}
&& H^{(2)}_{i}(\vec{k},\tau) =  \frac{(-i)^2}{(2\pi)^{3}}\int d^{3} q \, \int d^3 p\, \int d^{3} q^{\,\prime} \, \int d^3 p^{\, \prime} 
\,\,\, \delta^{(3)}(\vec{k} - \vec{q} - \vec{p})\,\,\, \delta^{(3)}(\vec{p} - \vec{q}^{\, \prime} - \vec{p}^{\, \prime})
 \nonumber\\
&&\times \int_{0}^{\tau} d\tau_{1} {\mathcal G}_{k}(\tau-\tau_{1})\, \, \int_{0}^{\tau_{1}} d\tau_{2} \, {\mathcal G}_{p}(\tau_{1}-\tau_{2})\,(q_{m} + p_{m}) \, (q_{m ' }' + p_{m' }') 
  \epsilon_{b\, m' \, n' \,} \, \epsilon_{a'\, b'\, n'} \, \epsilon_{m\, n\, i} \, \epsilon_{a\, b\, n}
 \nonumber\\
 &&\times
  \,  \langle v_{a'}(\vec{q}^{\,\,\prime},\tau_{2}) \, v_{a}(\vec{q},\tau_{1}) \rangle \,B_{b'}(\vec{p}^{\,\prime}).
\label{NM19}
\end{eqnarray}
After averaging the whole series of Eq. (\ref{NM14aa}) term by term the obtained result can be resummed and written as:
\begin{equation}
H_{i}(\vec{k},\tau) = \langle B_{i}^{(0)}(\vec{k},\tau)\rangle  + \langle B_{i}^{(2)}(\vec{k},\tau)\rangle + \langle B_{i}^{(4)}(\vec{k},\tau)\rangle +... =  e^{ - k^2 \, 
\overline{\nu}_{\mathrm{magn}}\, \tau} B_{i}(\vec{k}).
\label{C21}
\end{equation}
where the magnetic diffusivity 
coefficient $\nu_{\mathrm{magn}} = 1/(4\pi \sigma) $ has been renormalized as
\begin{equation}
\overline{\nu}_{\mathrm{magn}} = \nu_{\mathrm{magn}} + v_{0}, \qquad v_{0} = \frac{\tau_{\mathrm{c}}}{3}\int\, \frac{d k}{k} \, 
\overline{M}^2_{{\mathcal R}}(k,\tau_{*})\,{\mathcal P}_{\mathcal R}(k)\, \sin^2{(k/k_{*})} e^{ - 2 k^2/k_{\mathrm{d}}^2}, 
\label{IM7F}
\end{equation}
and $k_{*} = 1/r_{\mathrm{s}}(\tau_{*})$. The averaging suggested here has been 
explored long ago in the related context of acoustic turbulence  by Vainshtein and Zeldovich \cite{vain1} (see also \cite{vain2}). Prior to decoupling, however, both $\langle v^2 \rangle \propto {\mathcal A}_{\mathcal R} \ll 1$ and  $R_{\mathrm{kin}}\ll 1$. The resum indicated in Eq. (\ref{C21}) seems then to be more plausible in the present case than in the 
one of a kinetically turbulent plasma with strong inhomogeneities.
If the Markovian approximation is relaxed the velocity correlator becomes
\begin{equation}
\langle v_{i}(\vec{q},\tau_{1}) v_{j}(\vec{p},\tau_{2}) \rangle = \hat{q}_{i} \,\hat{q}_{j} \, \tilde{v}(q) \, \Gamma(q,\tau_{1},\tau_{2}) 
\delta^{(3)}(\vec{q} + \vec{p}), \qquad \tilde{v}(q) = \frac{2 \pi^2}{q^3} {\mathcal P}_{{\mathcal R}}(q).
\label{NM10}
\end{equation}
Assuming for sake of simplicity, $R_{\mathrm{b}}(\tau_{*})\ll 1$ and $c_{\mathrm{sb}}(\tau_{*}) \simeq 1/\sqrt{3}$ we have that 
 \begin{equation}
 \Gamma(q, \tau_{1},\tau_{2}) =\frac{3}{50} \biggl\{ \cos{[q c_{\mathrm{sb}} (\tau_{1} - \tau_{2})]} - 
 \cos{[q c_{\mathrm{sb}} (\tau_{1} + \tau_{2})]}\biggr\} e^{- q^2 \nu_{\mathrm{th}} ( \tau_{1} + \tau_{2})}.
\label{NM9}
\end{equation}
The results (\ref{IM7F}) and (\ref{NM9}) lead to physically equivalent estimates and support the conclusion
 that the diffusivity wavenumber is smaller than expected from the usual 
 arguments of the magnetic diffusivity scale at last scattering. In fact, ignoring 
 the contribution of the bulk velocity in Eq. (\ref{pd5})  the diffusivity wavenumber can be roughly estimated as $k_{\sigma} \sim \sqrt{{\mathcal H}_{*} \sigma}$. The 
 standard estimate of $k_{\sigma}$ must be compared with the diffusivity scale
 arising from the effect of the large-scale flow:
\begin{eqnarray}
&& k_{\sigma} \simeq 2.55 \times 10^{10} \biggl(\frac{d_{\mathrm{A}}}{14116 \, \, \mathrm{Mpc}} \biggr)^{-1/2}  \,\,\,\mathrm{Mpc}^{-1},
\nonumber\\
&& k_{v} \sim 50 \biggl(\frac{{\mathcal A}_{{\mathcal R}}}{2.41\times 10^{-9}}\biggr)^{- \frac{1}{n_{\mathrm{s}} +1}}\,
\biggl(\frac{d_{\mathrm{A}}}{14116 \, \, \mathrm{Mpc}} \biggr)^{-\frac{2}{n_{\mathrm{s}} +1}}\, \,\, \mathrm{Mpc}^{-1},
\label{NM11}
\end{eqnarray}
where $d_{\mathrm{A}}$ denotes the (comoving) angular diameter distance to last-scattering for the typical set 
of fiducial parameters of Eq. (\ref{R3a}).
The standard analysis based on the magnetic diffusivity scale would imply that all the modes $k > k_{\sigma}$ are 
diffused. The presence of large-scale flow implies that diffusion operates already for $k \geq k_{v}$. This means that the correct diffusion scale to be considered prior to decoupling is not $k_{\sigma} \sim {\mathcal O}(10^{11}) \mathrm{Mpc}^{-1}$ but, at most, $k_{v}\sim {\mathcal O}(50) \, \mathrm{Mpc}^{-1}$ which is closer to the Silk damping scale but qualitatively and quantitatively different.

In summary, prior to electron-positron annihilation the large-scale (turbulent) flow can only be determined indirectly from the features of the various phase transitions. After electron-positron annihilation the  flow can be 
inferred directly from the evolution of large-scale curvature perturbations imprinted in the CMB anisotropies.  The hierarchy between the kinetic and magnetic Reynolds numbers prior to last scattering suggests an effective description 
of the evolution of pre-decoupling magnetic which encompasses the conventional approach solely based on the conservation of the magnetic flux. On a more speculative ground the present considerations suggest that the  Reynolds numbers are somehow related to the largeness of the Hubble entropy during the early stages of the evolution of the plasma.

\end{document}